\documentclass[conference]{IEEEtran}
\IEEEoverridecommandlockouts
\usepackage{cite}
\usepackage{amsmath,amssymb,amsfonts}
\usepackage{algorithmic}
\usepackage{graphicx}
\usepackage{textcomp}
\usepackage{xcolor}
\usepackage{multirow}
\usepackage{siunitx}

\def\BibTeX{{\rm B\kern-.05em{\sc i\kern-.025em b}\kern-.08em
    T\kern-.1667em\lower.7ex\hbox{E}\kern-.125emX}}
\begin{document}

\title{Unit-Independent Low-Rate Wrist GSR Processing for Stress Detection Using Phasic nSCR Features\\

}
\author{
\IEEEauthorblockN{
Zequan Liang$^{1}$,
Sally Hang$^{2}$,
Geneva Jost$^{2}$,
Ning Miao$^{3}$,
Wei Shao$^{1}$,
Mahdi Pirayesh Shirazi Nejad$^{3}$,\\
Hossein Sayadi$^{4}$,
Ehsan Kourkchi$^{3}$,
Setareh Rafatirad$^{1}$,
Camelia E. Hostinar$^{2}$,
Houman Homayoun$^{3}$
} \\
\IEEEauthorblockA{
$^{1}$Department of Computer Science, University of California, Davis, Davis, CA, U.S.A.\\  
$^{2}$Department of Psychology, University of California, Davis, Davis, CA, U.S.A.\\ 
$^{3}$Department of Electrical and Computer Engineering, University of California, Davis, Davis, CA, U.S.A.\\ 
$^{4}$Department of Computer Engineering and Computer Science, California State University, Long Beach, Long Beach, CA, U.S.A.\\ 
Email: \{zqliang, salhang, gmjost, nmiao, wayshao, pirayesh\}@ucdavis.edu, \\ hossein.sayadi@csulb.edu, \{ekay, srafatirad, cehostinar, hhomayoun\}@ucdavis.edu} } 

\maketitle

\begin{abstract}

Galvanic skin response (GSR) is widely used for stress detection, but wrist-based GSR remains challenging because its absolute amplitude can differ substantially from laboratory-grade palmar measurements. In this paper, we propose a unit-independent low-rate wrist GSR processing pipeline to extract the number of skin conductance responses per minute (nSCR/min) as a stress-related feature. We collect paired wrist and palmar GSR recordings from 31 participants during sitting baseline, standing baseline, neutral speaking, and the Trier Social Stress Test (TSST), a laboratory social stressor task. The proposed pipeline cleans the raw GSR signal, decomposes it into tonic skin conductance level (SCL) and phasic skin conductance response (SCR), applies robust z-score normalization, and detects phasic SCR peaks to compute nSCR/min. Using random forest on 25Hz We-Be GSR, nSCR/min achieved balanced accuracies of 0.823 and 0.871 for binary classification between TSST and the sitting and standing baselines, respectively. Moreover, the 25Hz We-Be GSR features achieved comparable balanced accuracy to the original 100Hz features across the evaluated tasks. These results suggest the feasibility of low-rate, unit-independent wrist GSR processing for wearable stress detection.

\end{abstract}


\section{Introduction}
Galvanic skin response (GSR), also known as electrodermal activity (EDA), is a widely used physiological signal that reflects sweat gland activation through changes in skin conductance \cite{bakker2011s}. Because GSR has been shown to reflect sympathetic nervous system activity, it has been widely used for stress detection. Wrist-based wearable devices provide a practical and mobile platform for continuous GSR monitoring. However, wrist-based GSR remains challenging because traditional laboratory physiological monitoring systems, such as MindWare, usually measure skin conductance from the palm or fingers, where sweat gland density and sympathetic responses are generally stronger. Since sweat gland distribution and local skin properties differ across body regions, wrist-based GSR can show a substantially different absolute conductance range from palmar or finger-based GSR \cite{van2012emotional}.

\begin{figure}[t]
    \centering
    \includegraphics[width= \linewidth]{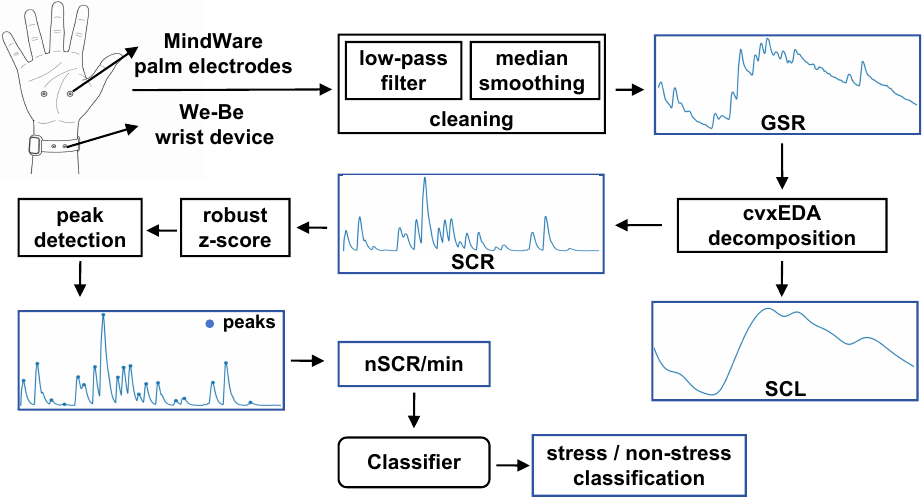}
    \caption{Overview of the GSR processing pipeline for stress detection}
    \label{fig:big_picture}
\end{figure}

To address this wrist-palmar amplitude mismatch, this study proposes a unit-independent processing pipeline for wrist-based wearable GSR stress detection, as shown in Fig.~\ref{fig:big_picture}. The pipeline cleans the raw wrist GSR signal, decomposes it into tonic skin conductance level (SCL) and phasic skin conductance response (SCR) components, applies robust z-score normalization, and detects phasic SCR peaks to extract the number of skin conductance responses per minute (nSCR/min) \cite{hu2024lab}. By combining normalization with phasic peak detection, the proposed pipeline shifts the analysis from conductance amplitude to phasic response rate, thereby capturing stress-related responses even under wrist-palmar measurement mismatch.

The contributions of this study are threefold: (1) We collect paired wrist and palmar GSR recordings from 31 adults across stress-related tasks to construct a dataset. (2) We propose a unit-independent wrist GSR processing pipeline to extract phasic nSCR features for stress detection, reducing the impact of wrist-palmar amplitude mismatch. (3) We show that 25Hz wrist GSR preserves segment-level stress detection performance compared with the original 100Hz data. These results support low-rate, unit-independent wrist GSR processing for wearable stress detection.

\section{Dataset Overview} 

\begin{figure}[t]
    \centering
    \includegraphics[width= \linewidth]{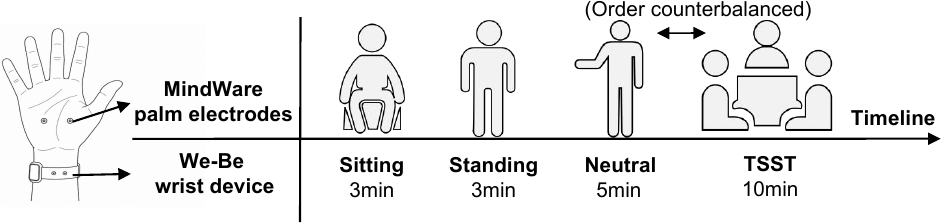}
    \caption{Timeline of the GSR data collection protocol}
    \label{fig:task_timeline}
\end{figure}

We collected GSR recordings from 31 adult participants. During each recording session, the We-Be wrist-worn device \cite{liang2025rapid} was placed on the participant's nondominant wrist using two electrodes on the inner side of the wristband. Meanwhile, the MindWare electrodes \cite{hu2024lab} were attached to the nondominant palm on the same side. This setup allowed simultaneous GSR acquisition from the wrist and palm of the same hand.

As shown in Fig.~\ref{fig:task_timeline}, each participant completed four task segments: a 3-minute sitting baseline, a 3-minute standing baseline, a neutral speaking task, and the Trier Social Stress Test (TSST). The last two speaking tasks were counterbalanced across participants to reduce carryover effects. During the 5-minute neutral speaking task, participants spoke to a video camera with no judges present about neutral topics, such as describing the room or explaining how to make a sandwich. The TSST was a 10-minute social stressor task~\cite{birkett2011trier}, in which participants performed a 5-minute job-candidacy speech followed by a 5-minute mental arithmetic task in front of two confederate judges.

After the completion of data collection, timestamps from the two devices were aligned, and the continuous GSR recordings were converted to the same skin conductance units (\si{\micro\siemens}) and divided into task-specific segments for further analysis.

\section{Data Preprocessing and Stress Detection} \label{sec:preprocessing_stress_detection}

As shown in the overview in Fig.~\ref{fig:big_picture}, after raw GSR data acquisition, the signals were processed through a GSR processing pipeline for stress detection. First, the raw GSR signal was cleaned using a 2Hz low-pass filter, followed by rolling median smoothing with a 1-second window to reduce high-frequency noise and small local fluctuations.

The cleaned GSR signal was decomposed using the NeuroKit2 implementation of cvxEDA, a convex-optimization-based method that models skin conductance as the sum of tonic, phasic, and residual components \cite{greco2015cvxeda}. The tonic component SCL captures slowly varying baseline activity, while the phasic component SCR captures rapid responses associated with discrete sympathetic arousal events \cite{braithwaite2013guide}. This separation allows us to focus on phasic SCR activity for stress detection while reducing dependence on absolute GSR amplitude.

Although the GSR, SCL, and SCR signals are measured in the same unit of skin conductance (\si{\micro\siemens}), their absolute amplitudes can differ substantially between palmar and wrist recordings due to the different measurement sites. This amplitude mismatch is particularly problematic for SCR peak detection, because conventional SCR detection often relies on fixed amplitude criteria, such as a minimum response amplitude of 0.05\si{\micro\siemens}. A threshold that is appropriate for palmar recordings may be too strict or too sensitive for wrist recordings. To reduce the effect of amplitude differences, we applied robust z-score normalization before SCR peak detection, as shown in Eq.~\ref{eq:robust_zscore}.

\begin{equation}
z_i = \frac{x_i - \mathrm{median}(x)}{1.4826 \times \mathrm{median}(|x_i - \mathrm{median}(x)|)}
\label{eq:robust_zscore}
\end{equation}

where $x_i$ represents the $i$th sample of the signal series $x$, and $z_i$ represents the corresponding robust z-score normalized value. 

After robust z-score normalization, SCR peaks were detected from the normalized phasic SCR signal. To reduce false detections caused by small local fluctuations, we applied a peak-detection criterion based on a minimum peak prominence of 0.2, a minimum peak width of 1 second, and a minimum distance of 1 second between adjacent peaks. The number of detected SCR peaks (nSCR) was then normalized by the duration of each task segment to obtain the number of SCRs per minute (nSCR/min).

For segment-level stress detection, we evaluated logistic regression, random forest, and a threshold-based classifier using nSCR/min as the input feature. Each model was used for binary classification between the TSST segment and non-stress segments.

\section{Experiment Results}

\subsection{Feature Statistics}

To evaluate low-rate wrist-based GSR processing, the original 100Hz We-Be GSR recordings were downsampled to 25Hz using anti-aliasing filtering, while the MindWare recordings were kept at their original 500Hz sampling rate. 

Table~\ref{tab:segment_feature_statistics} shows that raw GSR and SCL had substantially different absolute ranges between We-Be and MindWare, confirming the measurement-site mismatch between wrist and palmar GSR. In contrast, nSCR/min showed a more comparable numerical range and a consistent increasing trend from baseline tasks to TSST in both devices. nSCR/min also achieved higher Pearson correlation coefficient (PCC) than raw GSR and SCL in most tasks, especially during neutral speaking and TSST.

\begin{figure}[b]
    \centering
    \includegraphics[width=\linewidth]{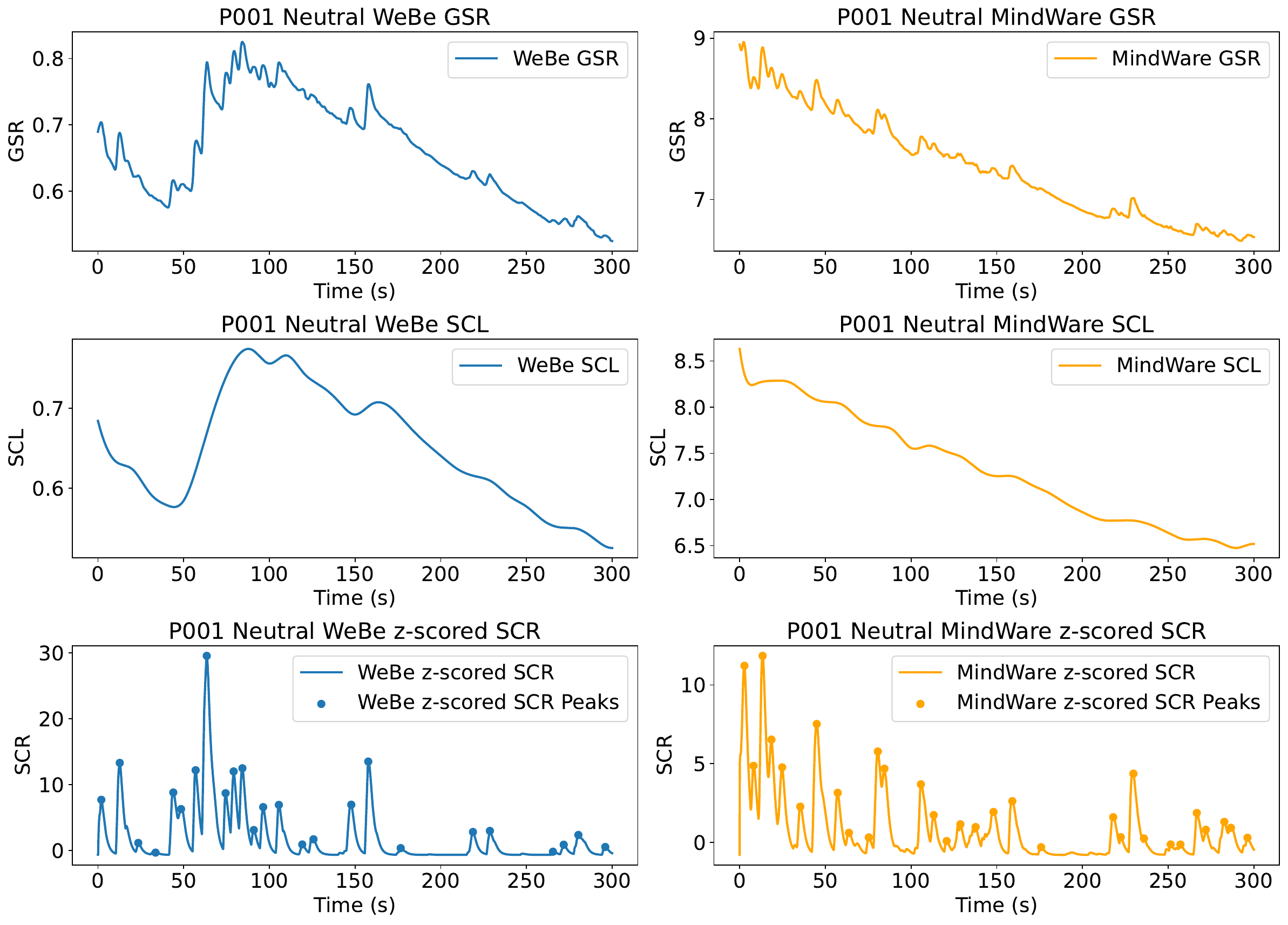}
    \caption{Representative GSR, SCL, and normalized SCR signals with detected SCR peaks}
    \label{fig:case_plot}
\end{figure}

Fig.~\ref{fig:case_plot} shows a representative case from one participant's task segment. The raw GSR and SCL signals also showed different absolute amplitude ranges between the two devices. After robust z-score normalization, both devices showed detectable phasic SCR peaks under the same normalized peak-detection criterion rather than a specific unit-based amplitude threshold.


\begin{table}[t]
\caption{GSR Feature Statistics and Correlations}
\setlength{\tabcolsep}{6.0pt}
\begin{center}
\begin{tabular}{|c|c|c|c|c|}
\hline
\textbf{Task} & \textbf{Feature} & \textbf{We-Be Mean} & \textbf{MindWare Mean} & \textbf{PCC} \\
\hline

\multirow{3}{*}{Sitting}
& GSR (\si{\micro\siemens}) & 0.031 & 4.868 & 0.012 \\
\cline{2-5}
& SCL (\si{\micro\siemens}) & 0.030 & 4.667 & 0.028 \\
\cline{2-5}
& nSCR/min & 3.557 & 5.482 & \textbf{0.193} \\
\hline

\multirow{3}{*}{Standing}
& GSR (\si{\micro\siemens}) & 0.033 & 6.096 & -0.006 \\
\cline{2-5}
& SCL (\si{\micro\siemens}) & 0.032 & 5.704 & \textbf{0.027} \\
\cline{2-5}
& nSCR/min & 3.735 & 5.695 & -0.048 \\
\hline

\multirow{3}{*}{Neutral}
& GSR (\si{\micro\siemens}) & 0.141 & 9.982 & 0.145 \\
\cline{2-5}
& SCL (\si{\micro\siemens}) & 0.137 & 9.458 & 0.148 \\
\cline{2-5}
& nSCR/min & 4.239 & 7.465 & \textbf{0.407} \\
\hline

\multirow{3}{*}{TSST}
& GSR (\si{\micro\siemens}) & 0.244 & 13.512 & -0.439 \\
\cline{2-5}
& SCL (\si{\micro\siemens}) & 0.227 & 12.578 & -0.396 \\
\cline{2-5}
& nSCR/min & 5.290 & 8.993 & \textbf{0.402} \\
\hline

\end{tabular}
\label{tab:segment_feature_statistics}
\end{center}
\end{table}

\subsection{Stress Detection}

Before conducting stress detection, we first evaluated whether nSCR/min showed task-level separation between the TSST and the non-stress conditions. Paired comparisons were performed between the TSST and each non-stress task, including sitting baseline, standing baseline, and neutral speaking. We also grouped all non-stress tasks together as a combined condition and compared it against the TSST.

\begin{table}[t]
\caption{Paired Comparison of nSCR/min Between TSST and Other Tasks}
\setlength{\tabcolsep}{4.0pt}
\begin{center}
\begin{tabular}{|c|c|c|c|c|}
\hline
\textbf{Task (TSST vs.)} & \textbf{Device} & \textbf{Mean Diff.} & \textbf{Wilcoxon $p$} & \textbf{Cohen $d_z$} \\
\hline

\multirow{2}{*}{Sitting}
& We-Be & 1.733 & $4.14\times10^{-4}$ & 0.779 \\
\cline{2-5}
& MindWare & 3.512 & $9.31\times10^{-10}$ & 1.659 \\
\hline

\multirow{2}{*}{Standing}
& We-Be & 1.555 & $3.86\times10^{-3}$ & 0.629 \\
\cline{2-5}
& MindWare & 3.299 & $8.20\times10^{-8}$ & 1.299 \\
\hline

\multirow{2}{*}{Neutral}
& We-Be & 1.033 & $5.99\times10^{-2}$ & 0.429 \\
\cline{2-5}
& MindWare & 1.486 & $1.04\times10^{-3}$ & 0.712 \\
\hline

\multirow{2}{*}{All Non-stress}
& We-Be & 1.468 & $1.59\times10^{-3}$ & 0.667 \\
\cline{2-5}
& MindWare & 2.756 & $1.77\times10^{-8}$ & 1.334 \\
\hline

\end{tabular}
\label{tab:paired_nscr_statistics}
\end{center}
\end{table}

Table~\ref{tab:paired_nscr_statistics} reports the paired statistical analysis of nSCR/min between the TSST and other tasks across subjects. The mean difference (Mean Diff.) was computed as the average paired difference between the TSST and the reference task across subjects. The Wilcoxon signed-rank test \cite{wilcoxon1945individual} was used to evaluate whether the paired differences between the TSST and each reference task were consistently nonzero across subjects, while Cohen's $d_z$ \cite{cohen1988statistical} was used to quantify the magnitude of these within-subject differences. 

Overall, MindWare showed larger mean differences than We-Be, indicating stronger task separation from the palmar reference device than from the wrist-based device. However, the We-Be wrist device still showed statistically significant discrimination between TSST and sitting baseline, standing baseline, and the combined non-stress tasks, with Wilcoxon $p<0.05$ and Cohen's $d_z>0.5$ in these comparisons. The weakest separation was observed for TSST versus neutral speaking, suggesting that the neutral speaking task may also elicit stress-related responses and is therefore more difficult to distinguish from the TSST.

For segment-level stress detection, we followed the same task settings as the paired statistical analysis and performed binary classification between the TSST task and each non-stress condition. We also included a combined setting in which all non-stress segments were grouped as the negative class. To avoid data leakage, all classification settings were evaluated using leave-one-subject-out (LOSO) cross-validation. 

Logistic regression, random forest, and a threshold-based classifier were evaluated using nSCR/min as the input feature. Logistic regression used within-fold feature standardization and class-balanced training, random forest was trained by 300 class-balanced trees, and the threshold was selected from training subjects by maximizing balanced accuracy. Performance was reported using balanced accuracy, AUC, and F1-score. Balanced accuracy was used as the primary metric to reduce the influence of class imbalance. AUC was not reported for the threshold-based classifier because it produced a fixed binary decision rather than a probabilistic output.

Tables~\ref{tab:webe_nscr_model_comparison} and~\ref{tab:mindware_nscr_model_comparison} show the stress detection results using We-Be and MindWare nSCR/min, respectively. For We-Be, random forest achieved the highest balanced accuracy for TSST versus sitting and standing, reaching 0.823 and 0.871, respectively. MindWare generally achieved higher performance than We-Be, consistent with stronger palmar electrodermal activity. Nevertheless, We-Be nSCR/min still provided discriminative information for detecting TSST-related stress. Across both devices, TSST was more distinguishable from sitting and standing baselines than from neutral speaking, which was the more challenging non-stress condition.

\begin{table}[t]
\caption{Stress Detection Performance Using \textbf{We-Be} nSCR/min Across Different Models}
\setlength{\tabcolsep}{3.0pt}
\begin{center}
\begin{tabular}{|c|c|c|c|c|}
\hline
\textbf{Task (TSST vs.)} & \textbf{Model} & \textbf{Balanced Accuracy} & \textbf{AUC} & \textbf{F1} \\
\hline

\multirow{3}{*}{Sitting}
& Logistic Regression & 0.758 & 0.688 & 0.706 \\
\cline{2-5}
& Random Forest & \textbf{0.823} & \textbf{0.824} & \textbf{0.825} \\
\cline{2-5}
& Threshold & 0.710 & -- & 0.654 \\
\hline

\multirow{3}{*}{Standing}
& Logistic Regression & 0.726 & 0.670 & 0.679 \\
\cline{2-5}
& Random Forest & \textbf{0.871} & \textbf{0.893} & \textbf{0.862} \\
\cline{2-5}
& Threshold & 0.742 & -- & 0.680 \\
\hline

\multirow{3}{*}{Neutral}
& Logistic Regression & \textbf{0.661} & 0.603 & \textbf{0.632} \\
\cline{2-5}
& Random Forest & 0.597 & \textbf{0.652} & 0.615 \\
\cline{2-5}
& Threshold & \textbf{0.661} & -- & 0.618 \\
\hline

\multirow{3}{*}{All Non-stress}
& Logistic Regression & \textbf{0.715} & 0.655 & 0.571 \\
\cline{2-5}
& Random Forest & 0.710 & \textbf{0.774} & 0.556 \\
\cline{2-5}
& Threshold & \textbf{0.715} & -- & \textbf{0.576} \\
\hline

\end{tabular}
\label{tab:webe_nscr_model_comparison}
\end{center}
\end{table}

\begin{table}[t]
\caption{Stress Detection Performance Using \textbf{MindWare} nSCR/min Across Different Models}
\setlength{\tabcolsep}{3.0pt}
\begin{center}
\begin{tabular}{|c|c|c|c|c|}
\hline
\textbf{Task (TSST vs.)} & \textbf{Model} & \textbf{Balanced Accuracy} & \textbf{AUC} & \textbf{F1} \\
\hline

\multirow{3}{*}{Sitting}
& Logistic Regression & \textbf{0.855} & \textbf{0.900} & 0.852 \\
\cline{2-5}
& Random Forest & 0.758 & 0.825 & 0.762 \\
\cline{2-5}
& Threshold & \textbf{0.855} & -- & \textbf{0.866} \\
\hline

\multirow{3}{*}{Standing}
& Logistic Regression & \textbf{0.839} & \textbf{0.889} & \textbf{0.839} \\
\cline{2-5}
& Random Forest & 0.742 & 0.780 & 0.733 \\
\cline{2-5}
& Threshold & 0.790 & -- & 0.800 \\
\hline

\multirow{3}{*}{Neutral}
& Logistic Regression & 0.613 & \textbf{0.718} & 0.613 \\
\cline{2-5}
& Random Forest & 0.532 & 0.543 & 0.525 \\
\cline{2-5}
& Threshold & \textbf{0.694} & -- & \textbf{0.732} \\
\hline

\multirow{3}{*}{All Non-stress}
& Logistic Regression & \textbf{0.801} & \textbf{0.838} & \textbf{0.658} \\
\cline{2-5}
& Random Forest & 0.629 & 0.698 & 0.451 \\
\cline{2-5}
& Threshold & \textbf{0.801} & -- & \textbf{0.658} \\
\hline

\end{tabular}
\label{tab:mindware_nscr_model_comparison}
\end{center}
\end{table}

\subsection{Sampling Rate}

We compared the stress detection performance on balanced accuracy between the original 100Hz GSR data and the downsampled 25Hz data from We-Be device, as shown in Table~\ref{tab:webe_25hz_100hz_balanced_accuracy}. The 25Hz We-Be GSR features achieved comparable or better performance than the 100Hz features across all tasks, suggesting that the lower sampling rate preserves sufficient information for segment-level stress detection.

Additionally, in our We-Be device, the GSR sensor is configured to use the same sampling rate as the PPG sensor. Since lower sampling rates can reduce sensing demands in wearable PPG systems~\cite{DBLP:conf/ndss/ShaoLZFMKRHF26}, these results suggest that a 25Hz configuration can preserve stress detection performance while supporting a lower-power wearable sensing design.

\begin{table}[t]
\caption{\textbf{Balanced Accuracy} Comparison Between 25Hz and 100Hz \textbf{We-Be} nSCR/min for Stress Detection}
\label{tab:webe_25hz_100hz_balanced_accuracy}
\setlength{\tabcolsep}{5.0pt}
\begin{center}
\begin{tabular}{|c|c|c|c|}
\hline
\textbf{Task (TSST vs.)} & \textbf{Model} & \textbf{25Hz} & \textbf{100Hz} \\
\hline

\multirow{3}{*}{Sitting}
& Logistic Regression & 0.758 & 0.726 \\
\cline{2-4}
& Random Forest & \textbf{0.823} & 0.645 \\
\cline{2-4}
& Threshold & 0.710 & 0.742 \\
\hline

\multirow{3}{*}{Standing}
& Logistic Regression & 0.726 & 0.661 \\
\cline{2-4}
& Random Forest & \textbf{0.871} & 0.823 \\
\cline{2-4}
& Threshold & 0.742 & 0.758 \\
\hline

\multirow{3}{*}{Neutral}
& Logistic Regression & \textbf{0.661} & 0.645 \\
\cline{2-4}
& Random Forest & 0.597 & 0.548 \\
\cline{2-4}
& Threshold & \textbf{0.661} & 0.565 \\
\hline

\multirow{3}{*}{All Non-stress}
& Logistic Regression & \textbf{0.715} & 0.683 \\
\cline{2-4}
& Random Forest & 0.710 & 0.575 \\
\cline{2-4}
& Threshold & \textbf{0.715} & 0.710 \\
\hline

\end{tabular}
\end{center}
\end{table}

\section{Discussion}
The results show that the number of SCR peaks captures stress-related phasic responses and enables the We-Be wrist device to detect stress patterns from low-rate GSR signals. Despite the amplitude differences between wrist-based and palmar GSR measurements, they showed consistent task-level trends in nSCR features, with the strongest discrimination observed between the TSST task and the sitting and standing tasks. Low-rate GSR sensing can also reduce power consumption, thereby benefiting battery life and supporting long-term continuous monitoring.

 Although the We-Be device achieved promising performance for segment-level stress detection, its performance was generally lower than that of the laboratory-grade MindWare system. This suggests that there is still room to improve stress detection accuracy on the We-Be device through better signal quality, preprocessing, and feature extraction in the future.

\section{Conclusion and Future Work}

This study addressed the challenge of wrist-based wearable stress detection, where absolute GSR and SCL amplitudes are sensitive to wrist-palmar measurement differences. To reduce this effect, we presented a unit-independent low-rate wrist GSR processing pipeline that combines robust normalization and phasic peak detection to extract nSCR/min as a response-rate feature rather than relying on absolute amplitude. Experimental results showed that low-rate We-Be GSR preserved discriminative information for binary classification between TSST-related stress and non-stress conditions. 

In future work, we will focus on improving We-Be signal quality, feature extraction, and motion-based noise removal using We-Be motion signals, to narrow the performance gap between wrist-based We-Be sensing and palmar laboratory-grade MindWare measurements.

\bibliographystyle{ieeetr}
\bibliography{EMBC}

@inproceedings{bakker2011s,
  title={What's your current stress level? Detection of stress patterns from GSR sensor data},
  author={Bakker, Jorn and Pechenizkiy, Mykola and Sidorova, Natalia},
  booktitle={2011 IEEE 11th international conference on data mining workshops},
  pages={573--580},
  year={2011},
  organization={IEEE}
}

@article{van2012emotional,
  title={Emotional sweating across the body: Comparing 16 different skin conductance measurement locations},
  author={van Dooren, Marieke and Janssen, Joris H and others},
  journal={Physiology \& behavior},
  volume={106},
  number={2},
  pages={298--304},
  year={2012},
  publisher={Elsevier}
}

@article{hu2024lab,
  title={From lab to life: Evaluating the reliability and validity of psychophysiological data from wearable devices in laboratory and ambulatory settings},
  author={Hu, Xin and Sgherza, Tanika R and Nothrup, Jessie B and Fresco, David M and Naragon-Gainey, Kristin and Bylsma, Lauren M},
  journal={Behavior Research Methods},
  volume={56},
  number={7},
  pages={1--20},
  year={2024},
  publisher={Springer}
}

@inproceedings{liang2025rapid,
  title={Rapid Adaptation of SpO2 Estimation to Wearable Devices via Transfer Learning on Low-Sampling-Rate PPG},
  author={Liang, Zequan and Zhang, Ruoyu and Shao, Wei and Kourkchi, Ehsan and Rafatirad, Setareh and Homayoun, Houman and others},
  booktitle={IEEE-EMBS International Conference on Body Sensor Networks 2025}
}

@article{greco2015cvxeda,
  title={cvxEDA: A convex optimization approach to electrodermal activity processing},
  author={Greco, Alberto and Valenza, Gaetano and Lanata, Antonio and Scilingo, Enzo Pasquale and Citi, Luca},
  journal={IEEE transactions on biomedical engineering},
  volume={63},
  number={4},
  pages={797--804},
  year={2015},
  publisher={IEEE}
}

@article{braithwaite2013guide,
  title={A guide for analysing electrodermal activity (EDA) \& skin conductance responses (SCRs) for psychological experiments},
  author={Braithwaite, Jason J and Watson, Derrick G and Jones, Robert and Rowe, Mickey},
  journal={Psychophysiology},
  volume={49},
  number={1},
  pages={1017--1034},
  year={2013}
}

@article{birkett2011trier,
  title={The Trier Social Stress Test protocol for inducing psychological stress},
  author={Birkett, Melissa A},
  journal={Journal of visualized experiments: JoVE},
  number={56},
  pages={3238},
  year={2011}
}

@inproceedings{DBLP:conf/ndss/ShaoLZFMKRHF26,
author = {Wei Shao and
Zequan Liang and
Ruoyu Zhang and
Ruijie Fang and
Ning Miao and
Ehsan Kourkchi and
Setareh Rafatirad and
Houman Homayoun and
Chongzhou Fang},
title = {Know Me by My Pulse: Toward Practical Continuous Authentication on
Wearable Devices via Wrist-Worn {PPG}},
booktitle = {33rd Annual Network and Distributed System Security Symposium, {NDSS}
2026, San Diego, California, USA, February 23-27, 2026},
publisher = {The Internet Society},
year = {2026},
url = {https://www.ndss-symposium.org/ndss-paper/know-me-by-my-pulse-toward-practical-continuous-authentication-on-wearable-devices-via-wrist-worn-ppg/},
bibsource = {dblp computer science bibliography, https://dblp.org}
}

@article{wilcoxon1945individual,
  title={Individual comparisons by ranking methods},
  author={Wilcoxon, Frank},
  journal={Biometrics bulletin},
  volume={1},
  number={6},
  pages={80--83},
  year={1945},
  publisher={JSTOR}
}

@article{cohen1988statistical,
  title={Statistical power analysis for the behavioral sciences New York},
  author={Cohen, Jacob},
  journal={NY: Academic},
  volume={54},
  pages={77--155},
  year={1988}
}

\end{document}